\documentclass[12pt,aps, pre]{revtex4}

\usepackage[english]{babel}
\usepackage[utf8]{inputenc}
\usepackage{amsmath,amssymb}
\usepackage{graphicx}

\begin{document}

\title{About Imperfect Mushroom Billiards}
\author{Karel Zapfe}
\affiliation{ICF-UNAM}
\email{karelz@fis.unam.mx}   
\author{Francois Leyvraz}
\affiliation{ICF-UNAM}
\email{leyvraz@fis.unam.mx}   
\author{Thomas H. Seligman}
\affiliation{ICF-UNAM}
\email{seligman@fis.unam.mx}   

\date{\today}

\bibliographystyle{plain}

\begin{abstract}

Imperfections of Bunimovich mushroom Billiards are
analyzed. Any experiment will be affected by such
imperfections, and it will be necessary to estimate
their influence. In particular some of
the corners will be rounded and
small deviations of the angle of the underside of the
mushroom head will be considered. The analysis displayed some
unexpected non-generic features.
The latter leads to a transition from a perfect mushroom
behavior to either
an ordinary KAM scenario or an abrupt transition to
complete chaos,
depending on the sign of the perturbation.
The former produces a fractal area of islands and
chaos, in fact a KAM scenario, not associated to
the
large island of stability of the mushroom billiard.

\end{abstract}

\maketitle

\section{Introduction}

Billiards have become the most physical or maybe the
least unphysical of examples favoured by
mathematicians and mathematical physicists to study
chaotic dynamics and dynamics at the transition form
order to chaos.
Other such examples include spaces of constant
negative curvature \cite{HopfNeg} and maps
\cite{Ruffo,Dana}.

Bunimovich's mushroom billiards \cite{Buhong} have
lately received much attention
\cite{Altmanmush, Altmanhamilton, Fried2007,Fredyo, bunimovreciente}.
They represent mixed systems which do not
show the generic fractal boundary between chaotic and
integrable
regions displayed e.g. by twist maps. The transition
between the
two regions is sharp, though some sticky parabolic
manifolds asociated to the boundary
have been found \cite{Altmanmush}.  The system is also
of interest because it displays
surprising patterns of return times to the foot of the
mushroom \cite{Fredyo}, and due to its
acessability to experiment \cite{Fried2007}. Yet
experiment implies lack of exactitude,
and this has caused us to study properties of
imperfect mushropom billiards.
Surprisingly we found, that the result is not only of
interest for error control,
but shows features that are interesting by themselves.

In the present paper we shall first recapitulate some
properties of mushroom billiards.  It was known since
their introduction
that they are highly non generic systems, and as such
we are interested in how far
from their normal behavior they drift under small
modifications. We will show the
significance of the departure from their perfect
versions under two classes of
simple modifications, which can be classified as
moving angles of straight walls or rounding
of corners. The second one can be regarded as a smoothing
on the billiard frontier.
We shall proceed to discuss the points in
this order and
illustrate the modifications in corresponding figures.
We shall discuss the results in general terms as well as their
implications for experiment.

\section{Bunimovich mushrooms}

\begin{figure}
\begin{center}
\includegraphics[width=0.9\textwidth]{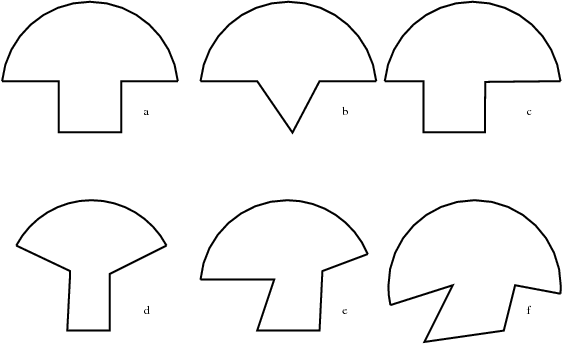}
\end{center}
\caption{Different perfect mushroom billiards. We
emphazise that in
\textbf{d},\textbf{e} and \textbf{f} the edges of the
straight walls of
the hat must point towards the center of the
circle.}\label{muchoshongos}
\end{figure}

A mushroom billiard is a classical planar billiard
with remarkable dynamics. It
is a Hamiltonian system in which the phase space
trivially decomposes into
an integrable and a chaotic part whose frontier is not
a fractal.
This billiard in it's simplest version is defined as
follows: Consider a boundary defined by
semi-circle and the diameter that closes this
semicircle.
Consider furthermore an opening in this diameter to
which the open side of some
rectangle or triangle is attached, to form three or
two more sides closing the boundary of the billiard.
If this rectangle or triangle does not touch the
circular part of the boundary,
we shall call the resulting billiard a
perfect mushroom (see fig. \ref{muchoshongos}). The
area  limited by the circle and the two
radial walls will be called the hat of the mushroom
and the rest the foot. It is not important that the
hat is truly a semi-circle, but it must be a sector of
a circle, meaning that its straight walls must point
exactly towards the center of the circle. The exact
shapes of the foot is irrelevant,
as long as it is starshaped, limited by straight lines
and not penetrating into the hat. Other options
use eliptic hats, but we shall not consider
these in the presnet paper.
The foot assures the effective free expansion of
bundles of rays, the mechanism with which we achieve
choas, i.e.
hyperbolicity on a subset of positive measure of phase space.
For a
exposition of this mechanism called defocusing,
see \cite{ChernMarkerg}.  Following these rules we can
create weird billiards that will
still have the most important properties of mushroom
billiards, see for example
the ones labeled \textbf{e} and \textbf{f} in the
figure \ref{muchoshongos}. For
simplicity we will consider only triangular or
rectangular feet and axially
symmetric mushrooms, as illustrated \textbf{a} and
\textbf{b} in fig.
\ref{muchoshongos}. Triangular feet allow us not to
worry about bouncing ball
orbits (parabolic manifolds) localized entirely inside the foot.

\begin{figure}
\begin{center}
\includegraphics[width=0.45\textwidth]{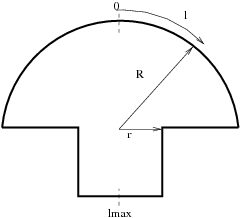}
\end{center}
\caption{Mushroom with the notation used.}\label{notation}
\end{figure}

We first introduce some notation as ilustrated in fig.
\ref{notation}.
The radius of the hat will be called $R$. $r$ will be
used for the distance from
the center of the circle to the farthermost inner
corner of the hat. For the Poincar\'e-Birkhoff
plots, we will take the arc length by the following
convention. The Zenith of the hat
will be the zero point, and it will increase to the
right along the border and decrease in the
opposite direction. As the border is topologically a
circle, we will use the
identification ofthe maximum and minimum arc lenght at
the point labeled $l_{max}$
(we are using only
symmetric mushrooms, for simplicity). Further notation
will be given as needed.

\begin{figure}
\begin{center}
\includegraphics[width=0.8\textwidth]{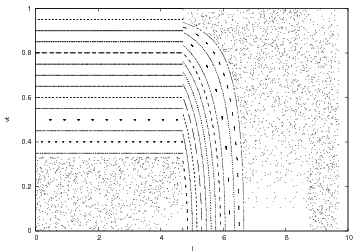}
\end{center}
\caption{Birkhoff map of the
flow for a perfect
symmetric mushroom. The Integrable part, with the
disconituities encloses the bounces of chaotic orbits
inside the hat.
The blank areas on the outside will be filled with
longer evolution, but acess is slow, as they also
contain the parabolic manifolds of the foot. Due to
time reversal invariance
and the axial symmetry of the
billiard, the section presents fourfold
symmetry. We will plot acordingly  only the positive cuadrant of the plot
.}\label{fasen}\end{figure}

The dynamic of the billiard is, as usual, given by
free motion inside the billiard
and specular reflection at the walls. Velocity is
normalized to the absolute
value $1$. As can be seen by image construction, the
orbits originating in the hat with angular momentum
larger than $r/R$ never leave the hat and are
completely integrable. Their caustic will define
a half circle whose radius is the absolute value of
the angular momentum,
a conserved quantity. Therefore $r$ is also the radius
of the smallest
possible caustic. Almost all other trajectories,
except subsets of measure zero, enter
the foot of the mushroom infinitely many times. There
they experience defocusing, and
are therefore chaotic. According to
Bunimovich's theorem \cite{Buteo}, they
define a subset of the phase space whose dynamic is
hyperbolic, ergodic and Bernoulli,
therefore chaotic in the usual sense. A
Poincar\'e-Birkhoff section illustrates this
(fig. \ref{fasen}).  A convenient parametrization of
the Poincar\'e-Birkhoff section is the arc
length of the border and the tangential (with respect
to this same border)  component
of the momentum of each trajectory. The map that
associates to each point in this
section the following bouncing point of the same
trajectory is called the Birkhoff map.
An example for a perfect mushroom is shown in figure
\ref{fasen}. Due to the lack of
differentiability at the corners of the billiard, some
curious, but not bothersome,
discontinuities appear in the integrable orbits. It
can be seen how the
conservation of the absolute value of angular momentum
produces a foliation in which
the bounces corresponding to the semicircle present a
constant tangential momentum. In the
following sections we will focus on this region of the
section. We will call
the integrable part $\mathfrak{I}$ and the chaotic
region $\mathfrak{P}$, both in the
flow and in the Birkhoff map.

\begin{figure}
\begin{center}
\includegraphics[width=0.8\textwidth]{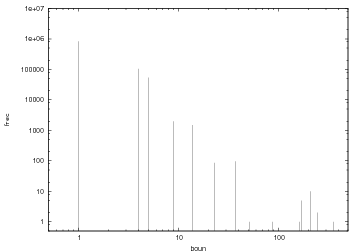}
\end{center}
\caption{Bouncing Frequency against number of bounces
for a perfect mushroom. Note the strong
selectivity in allowed bouncing times. The ratio of
the width of the opening against the radius
of the hat is $1/3$.}\label{maginun}
\end{figure}

Ref.\cite{Fredyo} presented a result concerning the
distribution of the allowed
numbers of bounces named \emph{magic
numbers} on the circlular part of the boundary between
any entry of the trajectory
into the hat and its next exit. For every ratio $r/R$
only some "magic" numbers are allowed and they can be
grouped,
according to their angular momentum, into
triads, in which the larger is sum of the other two.
This result is consequence of  a
theorem for rigid circle maps \cite{Slater}.
In figure \ref{maginun} this distribution is shown for
a perfect mushroom with aspect ratio $r/R=1/3$.

\section{Tilted Mushrooms}

We now consider a first type of perturbations:
We shall deal with deviations in the angle of the
straight lines forming the underside of
the hat with he circular boundary from the ideal value
of $\pi/2$. The tilting of
the underside of the hat will be shown on the right hand 
and measured in terms of the angle  $\epsilon$. The
other side of the picture refers
to a perturbation discussed below.

\begin{figure}
\begin{center}
\includegraphics[width=0.9\textwidth]{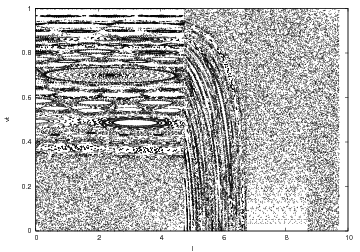}
\end{center}
\caption{An inverse tilted mushroom generates this
kind of phase space, a very typical KAM structure.
The tilting is $-0.2$ rad. Other tiltings generate
very similar pictures.}\label{opotilt}
\end{figure}

The angle to the circular border is no longer $\pi/2$.
and this leads to two different scenarios:

1) If this angle is smaller than $\pi/2$ i.e.
$\epsilon < 0 $ we find a transition to a mixed
phase space
with the usual charcateristics of a twist map. We
studied this case numerically and
it displays the generic scenario see figure
\ref{opotilt}. These billiards
can be considered as closely related to D billiards,
i.e. cicle sections,
where we have an abrupt transition from generic twist
map behaviour to total chaos passing through the
integrable half-circle.

\begin{figure}
\begin{center}
\includegraphics[width=0.9\textwidth]{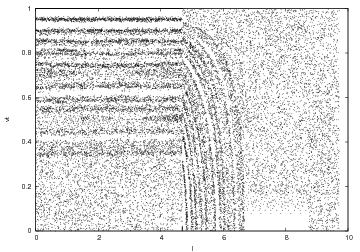}
\end{center}
\caption{Birkhoff-map of the
flow for a tilted mushroom.
The bouncing ball orbits still make the mixing rate in
their neighborhood very slow. Some integrable orbits
of the untilted mushroom are mimicked for short times,
but the covering will become uniform in time.   The
tilting parameter is $\epsilon=0.01$ radians. The map
has been iterated 500 times.}
\label{efergod}
\end{figure}

2) If this angle is larger than $\pi/2$ i.e. $\epsilon
> 0 $ we find ergodicity and indeed a
$K$-system. In this case the prolongation of the
edges,
that constitute the underside of the hat will
intersect at a point more distant than the center of the circle.
Allmost evry trajectory originating on the circle will eventually
hit these walls, wich are farther away than a  of the circle.
By image construction, this produces an efective free
expansion larger
than a corresponding cord of the circle, and then,
according to Bunimovich theorem \cite{Buteo}, a set of full measure
will have ergodic, mixing and hyperbolic dynamics.
Note that this does not exclude tye possible existence of
parabolic manifolds of measure zero.
Eventually, allmost all trayectories will leave the hat,
as they are in the ergodic
component of the phase space.   
The proof for the Kolmogorov property follows
\cite{bunimovreciente}. The map will present the apareance of
a blurred Birkhoff Map of the perfect mushroom in the short range 
(fig. \ref{efergod}).

\begin{figure}
\begin{center}
\includegraphics[width=0.8\textwidth]{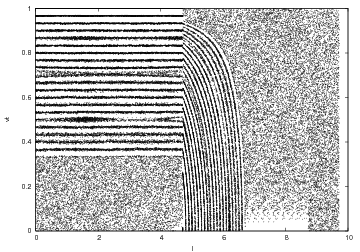}
\end{center}
\caption{The fundamental region where the shadow of
integrable orbits can be seen. The tilting
is $0.003$ radians and the map has been iterated
thousand times. It is important to observe the
slow mixing rate.}\label{zommtran}
\end{figure}

\begin{figure}
\begin{center}
\includegraphics[width=0.8\textwidth]{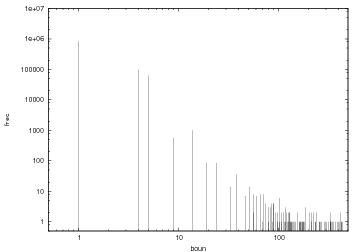}
\end{center}
\caption{Bouncing Frequency against number of bounces
for a tilted mushroom, $\epsilon=0.03$.
Compare with the plot \ref{maginun}. Selectivity is
lost at large bounce numbers.  \label{unmagnum}}
\end{figure}

If the tilting is very small (from the order of a few
thousandths of a radian), then,
for considerable times, the behavior of the billiard
shadows that of a perfect mushroom. The billiard is
now a generalized stadion but, nonetheless, dynamics
will be
unindistinguible from the perfect mushroom for some
period of time except for a
small though finite measure set of initial conditions. In
such {\it transient}
states, confusing structures do appear, as can be seen
in fig. \ref{zommtran}, near the point $(l,v_t)=(2,0.5)$. 
The arguments given in\cite{Fredyo} for the bouncing number 
distributions are no longer strictly valid, and the
selectivity of escape times is no longer rigid. Two
consecuences are to be expected: First
we will find low frequency ocurrences of small bounce
numbers outside the magic numbers,
as a consecuence of hitting the hole rather than the
hat by a small deviation. Second we
expect depeletion of the large magic numbers, as for
many bounces on the hat the angular deviation due to
the tilt
acumulates. This can also be viewed as a a consecuence
of the fact
that the parabolic manifolds near which the bounce
numbers diverge for the perfect
mushrooms no longer exist. We may
compare the figures \ref{maginun} and \ref{unmagnum},
in which the frequency of bounces
in the circle of the hat is plotted.
times, particularly at longer times.

\section{Rounded edges in Mushrooms}

\begin{figure}
\begin{center}
\includegraphics[width=0.9\textwidth]{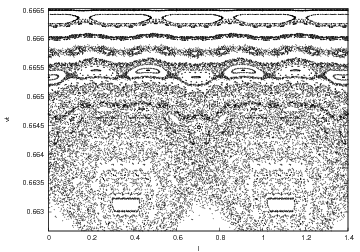}
\end{center}
\caption{A beautifull interwoven pattern of
KAM islands with centers of high period. The parameters are $\rho=1.0,
 \epsilon=0$. These islands are uncorrelated with the Larger island. 
They owe their existence  to a smoothing of the billiard \cite{bunimovreciente}.
 Notice the scale in both axis. The islands are invisible in the full plot.}
\label{islita01}
\end{figure}

The second deformation of the billiard that we are
going to consider is to round the inner edges
of the hat. We substitute the corners with defocusing
arcs of circles,
such that the curve representing
the boundary as well as its first derivative remain
continuous. This kind of deformations are called smoothings of
the billiard \cite{bunimovreciente}.
Obviously the second derivative
will not be continuous. The  caustic limiting the
integrable part has a larger radius, corresponding to
the point where the straight edge toward the circle
bounding the hat begins.
On the other hand the conditions requiered for chaos
in the remaining phase space
of the mushroom are now no longer fulfilled. We thus
expect a region with a mixed
phase space to appear and likely to be concentrated at
the edges of the integrable island which persists outside the larger caustic.
The large island must persist due to the fact that the angular momentum is still
conserved outside the innermost caustic.
Numerical studies confirm
this expectation, as can be seen in the figure \ref{islita01}.
The mechanism that creates this mixed phase space is
similar to the one found in eccentric ring billiards.

\section{Rounded and tilted mushrooms}

\begin{figure}
\begin{center}
\includegraphics[angle=-90,width=0.9\textwidth]{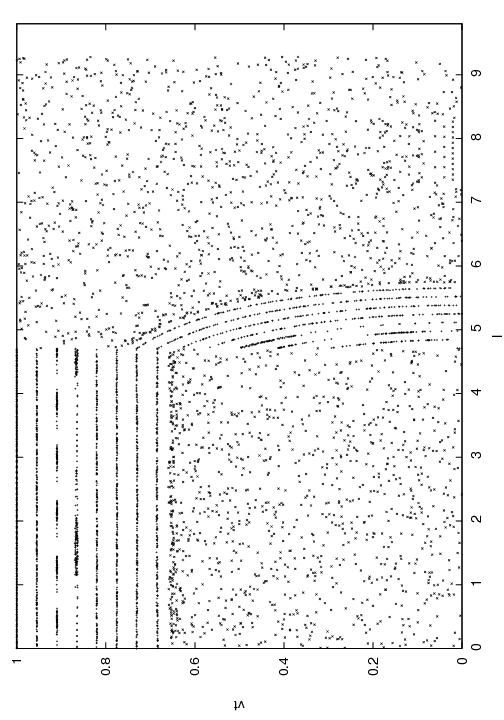}
\end{center}
\caption{Fundamental domain for the tilted and rounded
mushroom, with parameters $\rho=1.0,
\epsilon=0.00001$. Integrable islands are imposible to
spot, and the tilting is too small to
notice the breaking of the larger integrable effect.
Mixing effects are not
apreciable yet (500 iterations). Notice once more the smallness of the scale.}\label{redandtilt}
\end{figure}

\begin{figure}
\begin{center}
\includegraphics[width=0.9\textwidth]{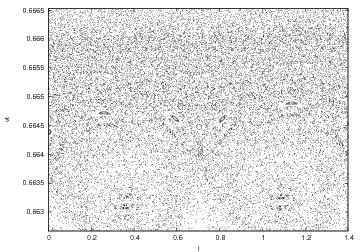}
\end{center}
\caption{A close-up of figure \ref{redandtilt}, which
shows the same region as
figure \ref{islita01}, but with tilting ($\epsilon=0.0002$).}
\label{tinyisland}
\end{figure}

If we consider the first kind of deformation, such
that it would
lead to complete chaos and combine it
with the second kind we get a picture, that without
the previous discussion would
seem very puzzeling.
The large integrable island, corresponding to the
conservation
of the magnitude of angular momentum, disappears, as
the tilt destabilizes the entire island.
For the curved surfaces on the other hand the scenario
is of KAM-type and
will move islands slightly and possibly change some
high level bifurcations, but not the
qualitative aspect of this mixed phase space area. It
becomes thus clear, that the
scenario we saw in the previous section is not of
typical KAM type for the
disolution of the mayor island. The mixed zone is truely
independent of the lagre stable island.
The latter becomes egodic for any tilting, thogh small
angles again cause longtransients.
The dispersing boundaries generate
a KAM structure not correlated to the disappearance of
the large island, i.e the mixed, fractal phase space region
is not qualitatively changed.
To see
this we show the Birkhoff
section of a mushroom with $\epsilon=0.0001$ and
$\rho=1.0d0$ in the figures \ref{redandtilt} and \ref{tinyisland}. The fractal
islands are very sensitive to
tilting, not because they have a direct correlation
with the disappareance of the larger islands, but
because even
a small angle of tilting will make a considerable perturbation at
the point of the rounding.
This billiard has no special
properties for long times, but shows the features of a
perfect mushroom billiard in transients.

\section{Conclusions and remarks}

We have seen that the characteristic behaviour of
mushroom billiards is non-generic with respect to
two possible one parameter perturbations. In the
case of tilting, the ``shadowing'' of the
integrable trajectories for fairly long times for
small values of $\epsilon$, making the orbits appear
as widened structures in the Birkhoff maps. The irruption of former
prohibited bouncing numbers in the bouncing
distribution of the hat can be understood
in terms of the breaking of the symmetry of the map \cite{Slater, Fredyo}. 

The central point of interest is the way in which
these billiards fail to be mushroom billiards.
In the case of the tilted mushrooms, they become, in
one direction, typical Bunimovich billiards
(hyperbolic, ergodic and Bernoulli), and in the other (inverse tilting),
generic KAM systems. Thus
the perfect mushroom billiards is a very peculiar
limiting case between a fractal with integrable
islands and global chaos, with just one clean
island. The time rates for the deviations
are also important, as they mark a limit for
acceptable precision during experiments of the type proposed in \cite{Fredyo}.  In the
case of rounded mushrooms, the prescence of KAM
islands could introduce noise in the experimental settings
of micro wave billiards. 
If they were noticiable, but as we have seen, this is
not the case.
The departures from the \emph{magic number} of bounces
distribution are also a point to be noticed.
As we could associate to each \emph{magic number} a
typical length for its path in the hat, it
could be devised a setting for measuring it
experimentally (\cite{Fried2007}). Imperfections of
the design could lead to the possible appearances of
other bouncing numbers than the magic ones,
and serve as a parameter for the cleanliness of the billiard
design, yet as we deal with wave expeiments, diffraction will hide any
small defects.

\section{Aknoledgements}

The authors would like to thank the following people and institutions.
Thomas Friedrich and Leonid Bunimovich for their usefull and insightefull discussions, CONACyT for the PhD schollarship, the CICCAC for lending us the facilities and resources, and the proyects IN-111607 (DGAPA-
UNAM) and 79988 (CONACyT), for the financial support.

\bibliography{hongobib}

\newpage

\begin{figure}
\begin{center}
\includegraphics[width=0.45\textwidth]{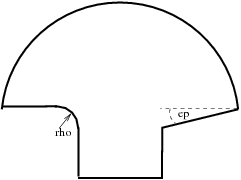}
\end{center}
\caption{Perturbed mushroom notation conventions.}\label{morenotation}
\end{figure}

\end{document}